%% file: Paper.tex
\pdfoutput=1
\documentclass[3p,times,number]{elsarticle}
\usepackage{lineno}
\usepackage{algpseudocode}
\usepackage{lipsum}
\usepackage{xcolor}
\usepackage{annotate-equations}
\usepackage{placeins}
\usepackage{amsmath,epsfig,subcaption,latexsym}
\usepackage{gensymb}
\usepackage{natbib}
\usepackage{mathtools,amssymb}
\biboptions{sort&compress}
\usepackage[justification=centering]{caption}
\usepackage{bm}
\usepackage{hyperref}
\usepackage{url}
\usepackage[export]{adjustbox}
\usepackage{textcomp}
\usepackage{array, tabularx,boldline}
\usepackage{cellspace}
\usepackage{xspace}

\makeatletter
\newcommand{\vast}{\bBigg@{4}}
\newcommand{\Vast}{\bBigg@{5}}
\makeatother
\long\def\symbolfootnote[#1]#2{\begingroup%
\def\thefootnote{\fnsymbol{footnote}}\footnote[#1]{#2}\endgroup}

\newcount\ndots
\def\drwln#1#2{\raise 2.5pt\vbox{\hrule width #1pt height #2pt}}

\def\square   {${\vcenter{\hrule height .4pt
               \hbox{\vrule width .4pt height 3pt \kern 3pt
               \vrule width .4pt}
               \hrule height .4pt}}$\nobreak\ }

\def\filsqr   {${\vcenter{\hrule height 2pt
                        \hbox{\vrule width 2.2pt height 0.2pt \kern 0.1pt
                              \vrule width 2.2pt}
                              \hrule height 2.2pt}}$\nobreak\ }
\usepackage{multicol}
\usepackage{dirtree}
\usepackage[frozencache,cachedir=minted-cache]{minted}
\usepackage{xcolor}
\newcommand\mydirtree[1]{{\small\dirtree{#1}\par}}
\newcommand\codetext[1]{{\small\texttt{#1}}}
\newcommand\codetextsmall[1]{{\scriptsize\texttt{#1}}}
\colorlet{fillcolor}{white!90!blue}
\newcommand\highlight[1]{\fcolorbox{white}{fillcolor}{#1}}%
\begin{document}
\begin{frontmatter} 
\title{Composable Design of Multiphase Fluid Dynamics Solvers in Flash-X}

\author[MCS]{Akash Dhruv \corref{cor2}}
\ead{adhruv@anl.gov} 



\address[MCS]{Mathematics and Computer Science Division,\\
Argonne National Laboratory, Lemont, IL, USA}


\cortext[cor2]{Corresponding author}
%

\begin{abstract}
Multiphysics incompressible fluid dynamics simulations play a crucial role in understanding intricate behaviors of many complex engineering systems that involve interactions between solids, fluids, and various phases like liquid and gas. Numerical modeling of these interactions has generated significant research interest in recent decades and has led to the development of open source simulation tools and commercial software products targeting specific applications or general problem classes in computational fluid dynamics. As the demand increases for these simulations to adapt to platform heterogeneity, ensure composability between different physics models, and effectively utilize inheritance within partial differentiation systems, a fundamental reconsideration of numerical solver design becomes imperative. The discussion presented in this paper emphasizes the importance of these considerations and introduces the Flash-X approach as a potential solution. The software design strategies outlined in the article serve as a guide for Flash-X developers, providing insights into complexities associated with performance portability, composability, and sustainable development. These strategies provide a foundation for improving design of both new and existing simulation tools grappling with these challenges. By incorporating the principles outlined in the Flash-X approach, engineers and researchers can enhance the adaptability, efficiency, and overall effectiveness of their numerical solvers in the ever-evolving field of multiphysics simulations.
\end{abstract}
%
\end{frontmatter}

\input{Tex/Introduction}
\input{Tex/Body}
\input{Tex/Conclusion}

\bibliographystyle{acm}
\bibliography{References/References}

\end{document}

%% file: Tex/Introduction.tex
%
%
\section{Introduction} \label{sc: introduction}
The recent advancements in High Performance Computing (HPC) have significantly accelerated progress in multiscale fluid dynamics simulations. Previously, simulations for phase-change heat transfer, which took a week on leadership supercomputers, can now be completed within hours, thanks to cutting-edge libraries developed under the Exascale Computing Project (ECP) \cite{DUBEY2022,AMReX_JOSS,Dubey2014,Oneal2018}. As we approach an era dominated by Artificial Intelligence (AI), Machine Learning (ML), and remarkable developments in specialized software/hardware infrastructure, it is imperative that simulation tools transition from a monolithic software architecture (where all features are packed into one source code) to a more flexible approach. This involves leveraging and combining different libraries to address complex and intriguing problems, and adopting a more strategic approach towards software development that exploits inheritance and design patterns within the mathematical system. 

Computational Fluid Dynamics (CFD), which is at the forefront of computational mechanics in utilizing large-scale computational resources to tackle problems of increasing size and complexity, can greatly benefited from this design philosophy. Direct Numerical Simulation (DNS), for example, which are high-fidelity, three-dimensional, time-dependent computations resolving all scales of motion down to the Kolmogorov scale, have provided valuable information on the fundamental physics of near wall turbulence \cite{lee_moser_2015}, transition \cite{WuE5292}, natural convection \cite{VerziccoRB2017}, and noise reduction in free shear flows \cite{kim_bodony_freund_2014}, just to name a few. In most cases, to enhance the range of scales that is captured (i.e. achieve Reynolds Number (Re) that is as close as possible to practical applications), specialized in-house solvers tuned to the particular configuration are utilized. These tools cannot be directly extended to more complex configurations of practical interest. In addition, rapid changes in hardware architectures require constant development/optimization of these solvers, which is usually beyond the capabilities of small, single Principal Investigator (PI) groups.

\begin{figure}[h]
\begin{center}
\includegraphics[width=0.3\textwidth]{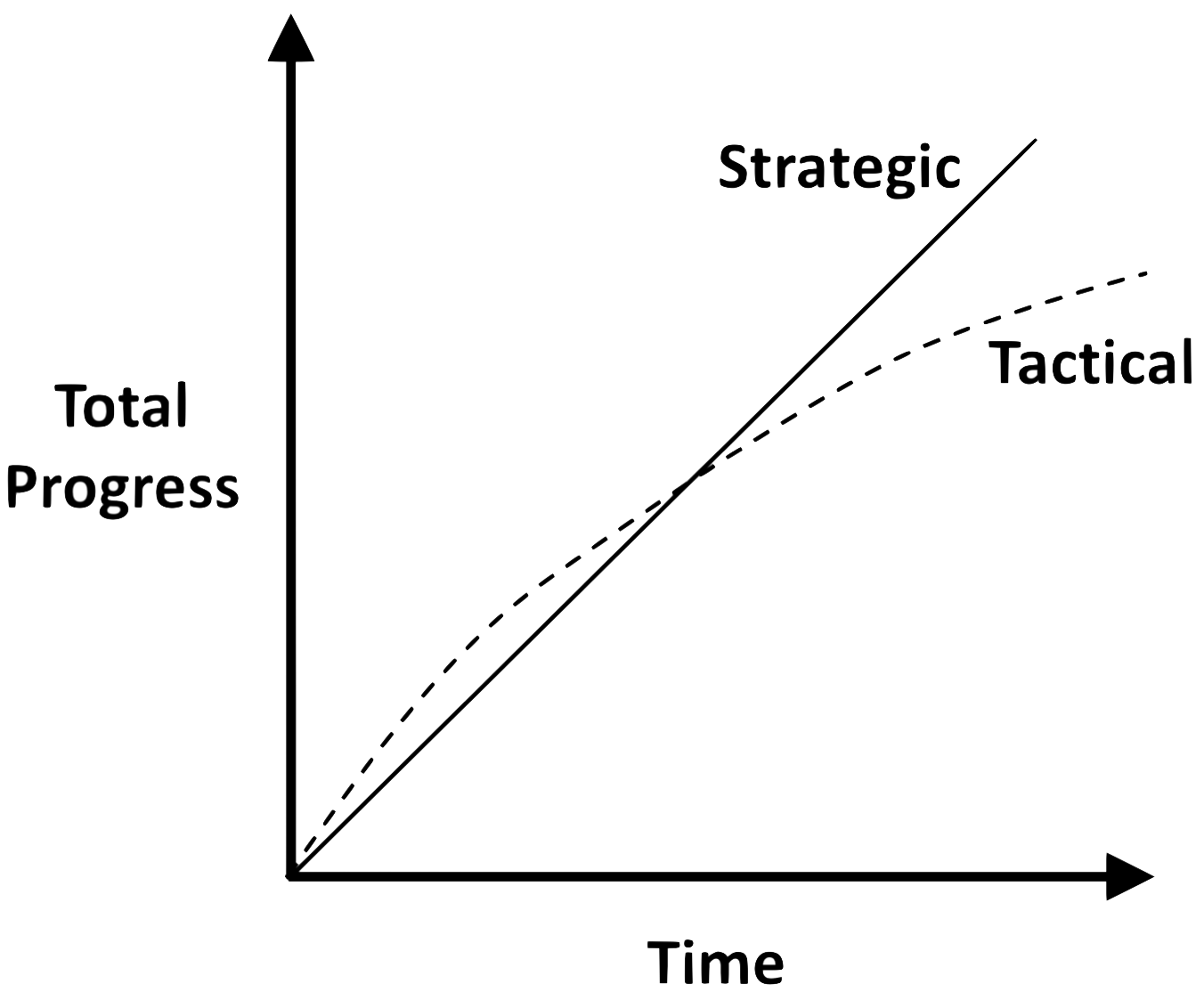}
\caption{Comparison of progress over time using strategic versus tactical programming approach. Tactical approach improves productivity in the beginning, but over time it introduces complexity within the software architecture that slows down progress. Using an agile programming methodology and following the strategic approach improves productivity over time. The figure is inspired from \cite{ousterhout2018philosophy}\label{fig:compare_software}.}
\end{center}
\end{figure}

A common problem with legacy CFD codes passed down from PIs to their students is the complexity of software design. This complexity is a symptom of inherent issues associated with waterfall programming model that has largely been the norm for university research laboratories, focusing on functionality of the code rather than software design. This typically starts at inception, where a student or a PI makes software design decisions related to a specific problem they are trying to simulate and freeze the code structure, not allowing any flexibility. When it is time to add more functionality to simulate a different problem, new software developments are implemented in form of patches around the original design contributing towards unsustainability. This approach is known as tactical programming and leads to an explosion of complexity. Agile, or strategic programming, on the other hand, simplifies design complexity by breaking down the software into small units, commonly referred to as modules. Each module encompasses/hides a complex software implementation, and can be used as a building block for a larger application. The focus is on developing a clean design where new developments can be easily implemented in the form of new modules, improving scalability and extensibility. Fig \ref{fig:compare_software} shows a comparison between the strategic and tactical approach that highlights the benefits of strategic programming over time \cite{ousterhout2018philosophy}.

Poorly designed research code can impede productivity for both students and experienced researchers, consuming valuable time as they navigate the intricacies of the code rather than efficiently applying or extending it to different scientific inquiries. The lack of modularity, inadequate documentation, absence of version control, inconsistent code style, limited testing, platform dependencies, inefficient algorithms, and inadequate error handling collectively contribute to this hindrance, underscoring the importance of prioritizing good software engineering practices to enhance code adaptability and overall usability across various use cases and hardware platforms. It is not surprising that research software engineering is increasingly becoming an important skill for any aspiring research professional in the field of computational engineering.

The technology industry has long embraced agile software design principles, drawing on the extensive expertise of software engineering teams. In scientific computing research, the need for agile design arises as PIs collaborate and merge their codebases. This collaboration often requires a complete redesign, and within this context, an open-source philosophy provides significant advantages, accelerating and scaling research. If a group of PIs, along with software development experts from national laboratories, collaborates on the same codebase, they can replicate the productivity evident in the technology industry.

Scalable, open-source CFD solvers with a broader applicability have been steadily gaining traction within the fluid dynamics community. Nek5000 \cite{Nek5000}, a spectral element solver designed for intricate geometrical configurations, serves as an exemplary case. It has been extensively applied to address turbulent and transitional fluid flow and heat transfer problems across diverse areas, from thermal hydraulics \cite{Merzari19} to wind energy \cite{10.1115/IMECE2015-53671}. Another instance is PyFR, an open-source Python-based framework for streaming architectures applicable to compressible flows \cite{PyFR}. It is crafted for mixed unstructured grids containing various element types and targets a spectrum of hardware platforms through the use of an in-built domain-specific language derived from the Mako templating engine.

The widespread adoption of solvers, including those mentioned above, as well as others like OpenFOAM \cite{OpenFOAM}, Phasta \cite{bolotnov2010,NAGRATH20054565,RODRIGUEZ2013115}, and Fire Dynamics Simulator (FDS) \cite{VanellaFDS}, signifies a paradigm shift within the broader fluid dynamics community. Despite the growing popularity, this community has been hesitant to embrace open-source development compared to other fields. This hesitancy stems from concerns related to composability, reproducibility, software quality maintenance, and complexities within the software infrastructure, making it challenging for them to extend to new applications and algorithms. The Flash-X team has dedicated substantial efforts over the past three years to tackle these challenges and enhance the robustness of the ecosystem for scientific software development and use. This article serves as an overview of this intricate system, offering a starting point for future Flash-X developers to familiarize themselves with the tools and practices established over time.

Many physics and engineering problems of interest involve interfaces between different types of fluids, such as gases and liquids, coexisting with complex geometrical configurations and moving or deforming solid boundaries. Effectively capturing these diverse interactions within an application demands a rigorous software development approach, an aspect that has been lacking in the field of scientific computing. FLASH, a community code for fluid dynamics, astrophysics, and magnetohydrodynamics simulations \cite{Dubey2012}, has played a crucial role in this field for the past 24 years. It laid the groundwork for a versatile computational software framework that seamlessly integrates Adaptive Mesh Refinement (AMR) with various Partial Differential Equation (PDE)-based physics problems. Over time, FLASH has evolved to encompass a broader range of applications. It was expanded to model single-phase fluid-structure interaction \cite{Vanella2009, Vanella2010} and multiphase pool boiling simulations \cite{DHRUV2019, DHRUV2021}. Recent developments have led to the creation of Flash-X \cite{DUBEY2022}, involving a substantial redesign to enhance performance and ensure compatibility with modern hardware architectures \cite{DUBEY2023,COUCH2021102830}. This showcases Flash-X's adaptability and commitment to remaining pertinent in the ever-evolving landscape of scientific computing.

The discussion presented in this work centers on the software design of a subset of Flash-X applications dedicated to modeling engineering problems involving interactions between gas, liquid, and moving/stationary solids \cite{akash_phd_2021}. Examples encompass a range of scenarios, including convective air-cooling of electronic devices \cite{YOUNG1998}, boiling-based cooling of machine components \cite{KIM20023919}, air-water dynamics during the breaking of surface waves \cite{farshad2019}, solidification-melt dynamics in metal casting \cite{griffin1991ladle}, liquid-gas chemical reactions in carbon capture \cite{holmes2012}, cold plasma-based treatment of cancer cells \cite{lgmartinez2019}, and more. In these applications, the interplay between interface dynamics and surrounding fluid motion is intricate. Factors such as density and viscosity ratios, temperature jumps across the interface, surface-tension effects, and boundary conditions significantly influence the dynamics. These complexities are captured through interface-resolving DNS, relying on the implicit representation of the interface, whose motion is tracked using a level-set function. We intentionally avoid delving into the mathematical and numerical formulation, providing references when necessary. Instead, our focus is on outlining details of the object-oriented design of physics and infrastructure units that leverages directory-based inheritance within the source code to allow for composability and extensibility.

The paper is structured as follows: Section \ref{sc: multiphase-system} provides and overview of the incompresible multiphase fluid dynamics implementation in Flash-X along with example applications. Section \ref{sc: software-design} focuses on details of the software design to serve as a guideline for Flash-X users and developers. Section \ref{sc: testing-reproducibility} focuses on tools and practices that we been developed around Flash-X for software maintenance and reproduciblity and finally we provide conclusion and the direction of future work.

%% file: Tex/Body.tex
%
%
\section{Multiphase Fluid Dynamics System in Flash-X} \label{sc: multiphase-system}
\begin{figure}[h]
\begin{center}
\includegraphics[width=0.87\textwidth]{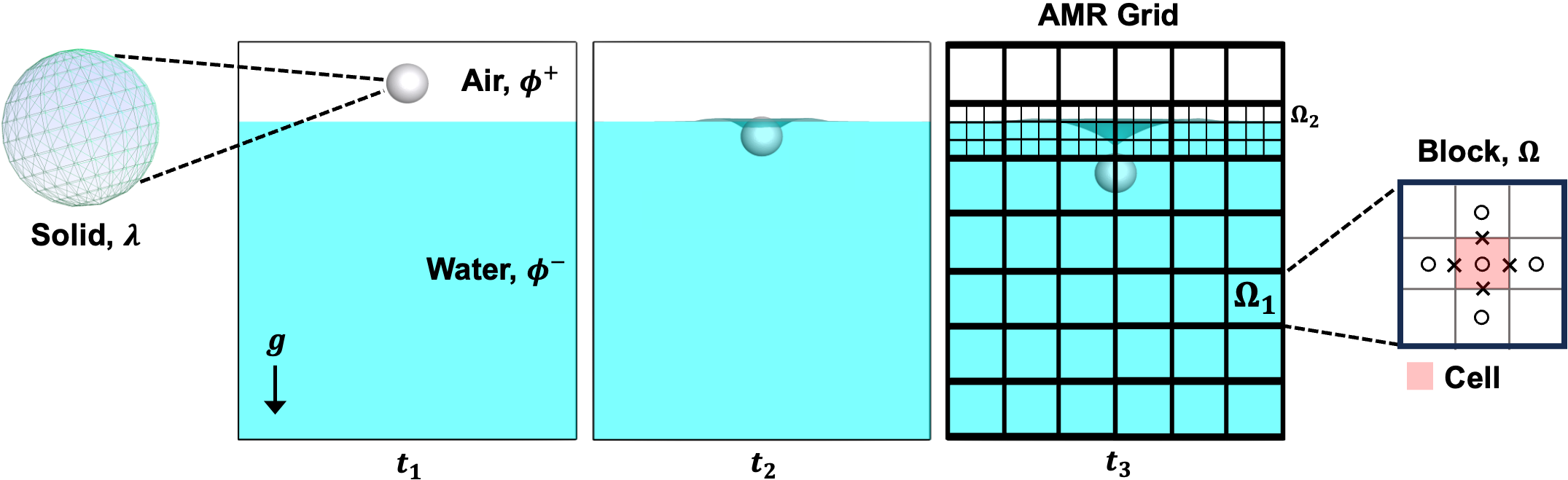}
\caption{Schematic of an example problem in Flash-X where a solid sphere, $\lambda$, disturbs the air-water free surface from $t_1$ to $t_3$ as it falls under the influence of gravity, $g$. The sphere is represented using a collection of Lagrangian triangles, and the air-water interface is defined using a signed level-set distance function, $\phi$, which is greater than zero in air ($\phi^+$) and less than zero in water ($\phi^-$). An example of Adaptive Mesh Refinement (AMR) is shown at $t_3$. AMR is applied at $\phi=0$, which corresponds to the implicit location of the interface. $\Omega_1$ and $\Omega_2$ represent two different resolutions for a block, $\Omega$, which contains a collection of cells with a staggered arragement of Cartesian points on faces ($\times$) and cell-centers ($\bullet$). This collection is knows as a Stencil, $S$ where partial differentiation equations are solved.\label{fig:grid}}
\end{center}
\end{figure}

Flash-X employs a staggered-grid, finite-difference, fractional-step projection method to solve incompressible fluid dynamics equations. These equations, derived from the Reynolds transport theorem for momentum, mass, and energy, are applicable to single-phase, multiphase, and phase-change heat transfer problems. The computations take place on a block-structured Cartesian grid, where each block $\Omega$ is treated as a data structure within the software design context. A block is composed of cells, further divided into points across three spatial dimensions. Partial differential equations are solved over time and space within these cells. The schematic in Figure \ref{fig:grid} illustrates a multiphase computational scenario, where a solid sphere $\lambda$ falls under gravity $g$, disturbing the free surface between air and water from time $t_1$ to $t_3$. $\lambda$ consists of triangular nodes treated as separate data structures. $\Omega$ and $\lambda$ represent distinct spatial reference frames, Eulerian and Lagrangian, respectively. Blocks $\Omega$ can exist at different resolutions, as demonstrated at time $t_3$ in Figure \ref{fig:grid}, where the resolution of $\Omega_2$ is higher than that of $\Omega_1$, implying smaller cell sizes in the former. Different resolutions within an Adaptive Mesh Refinement (AMR) grid enhance the accuracy of numerical solutions, particularly in regions of interest like the air-water interface. This interface is represented using a signed level-set distance function $\phi$, which is positive in air, negative in liquid, and zero at the implicit location of the interface.
\begin{figure}[h]
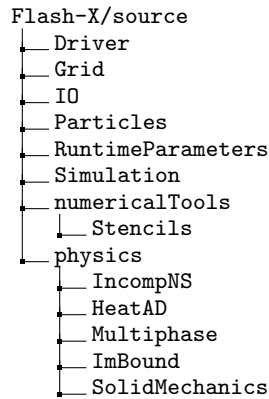

\begin{center}
\begin{subfigure}[b]{0.25\textwidth}
\mydirtree{%
.1 Flash-X/source.
.2 Driver.
.2 Grid.
.2 IO.
.2 Particles.
.2 RuntimeParameters.
.2 Simulation.
.2 numericalTools.
.3 Stencils.
.2 physics.
.3 IncompNS.
.3 HeatAD.
.3 Multiphase.
.3 ImBound.
.3 SolidMechanics.
}
\end{subfigure}
\end{center}
\caption{Top-level overview of Flash-X source code for multiphase fluid dynamics applications.}
\label{fig:source-overview}
\end{figure}

An essential design pattern to highlight is the stencil, denoted as $S$. It is characterized by a staggered arrangement of face ($\times$) and cell ($\bullet$) center points. Vector quantities are positioned at face centers, while scalars are situated at cell centers. Consider the following example of a partial differentiation equation representing the transport of a physical quantity, $Q$ (which can be either a vector or a scalar), through space and time,

\begin{equation} \label{eq:transport}
\begin{split}
\eqnmarkbox[magenta]{time}{\frac{\partial Q}{\partial t}} - \eqnmarkbox[teal]{advection}{\vec{u} \cdot \nabla{Q}} = \eqnmarkbox[red]{diffusion}{K \nabla^2 Q} + \eqnmarkbox[blue]{forcing}{f}
\end{split}
\end{equation}
\annotate[yshift=0.5em,color=black]{above,left}{time}{Temporal Evolution}
\annotate[yshift=-0.05em,color=black]{below,right}{advection}{Advection}
\annotate[yshift=0.5em,color=black]{above,right}{diffusion}{Diffusion}
\annotate[yshift=-0.25em,color=black]{below,right}{forcing}{Source Terms}

Equation \ref{eq:transport} is derived from the Reynolds transport theorem and identifies various continuum forces acting on $Q$. Temporal evolution denotes changes over time. Advection forces arise from the transport induced by the velocity $u$, while the diffusion force is attributed to variations in the concentration of $Q$ over space and is a multiple of the diffusion coefficient $K$. The term $f$ signifies the impact of additional source terms, such as gravity, pressure gradients, and other body forces. The operators for each force can be generalized over the stencil, $S$, and become separate design patterns. Effective software design can be achieved by exploiting these design patterns and the data structures $\lambda$ and $\Omega$ discussed above.
\begin{figure}[h]
\begin{center}
\includegraphics[width=0.8\textwidth]{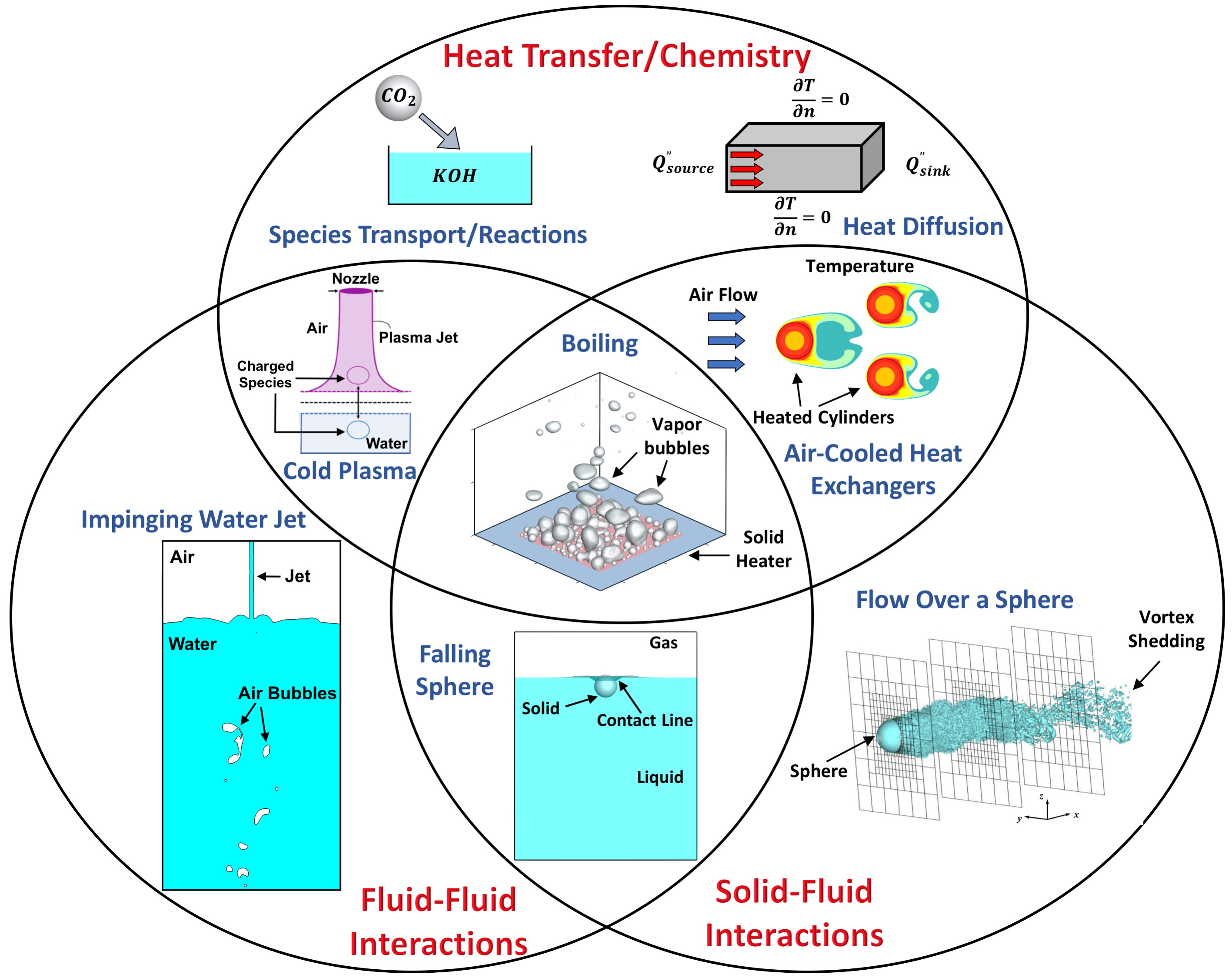}
\caption{Venn diagram of multiphysics categories that describe interactions involved in practical engineering problems\label{fig:overview}}
\end{center}
\end{figure}

The Flash-X source code organization provides a solid foundation for incorporating software design elements for $\lambda$, $\Omega$, and $S$. The codebase is derived from FLASH which was extensively developed by Vanella \cite{Vanella2009,Vanella2010}, Delaney \cite{Daley2012}, Wang \cite{WANG2019240} for multiphase and fluid-structure interaction problems. While the design of the source code will be discussed at length in the following sections, a top-level overview is provided in Figure \ref{fig:source-overview}. Directories under \codetext{Flash-X/source} are termed as units, each containing code pertinent to its functionality. For instance, the \codetext{Grid} and \codetext{Particles} units contain code relevant to the management of $\Omega$ and $\lambda$, respectively. The \codetext{Stencils} unit under \codetext{numericalTools} manages various stencils related to advection, diffusion, and temporal integration in Equation \ref{eq:transport}. The \codetext{Driver} unit manages the parallel execution environment and the integration/evolution of various physics components, while the \codetext{IO} oversees read and write functionality for data. Definition and management of execution time parameters is handled by \codetext{RuntimeParameters} unit.

The \codetext{physics} units contains specific algorithms for solving various mathematical models,
\begin{itemize}
    \item \codetext{IncompNS}: Routines for solving incompressible Navier-Stokes equations, both with constant and variable coefficients, to compute solutions for velocity and pressure.
    \item \codetext{HeatAD}: Routines for solving heat diffusion and advection equations, accommodating both constant and variable coefficients, to compute solutions for temperature.
    \item \codetext{Multiphase}: Routines for handling multiphase fluid properties and boundary conditions on physical variables, with and without evaporation. These routines are designed for solving the level-set function $\phi$ and setting source terms for velocity and temperature.
    \item \codetext{ImBound}: Modeling effects of the solid, $\lambda$, on continuum fluid.
    \item \codetext{SolidMechanics}: Modeling effects of fluid and other body forces on $\lambda$.
\end{itemize}
\begin{figure}[h]
\begin{center}
\includegraphics[width=0.88\textwidth]{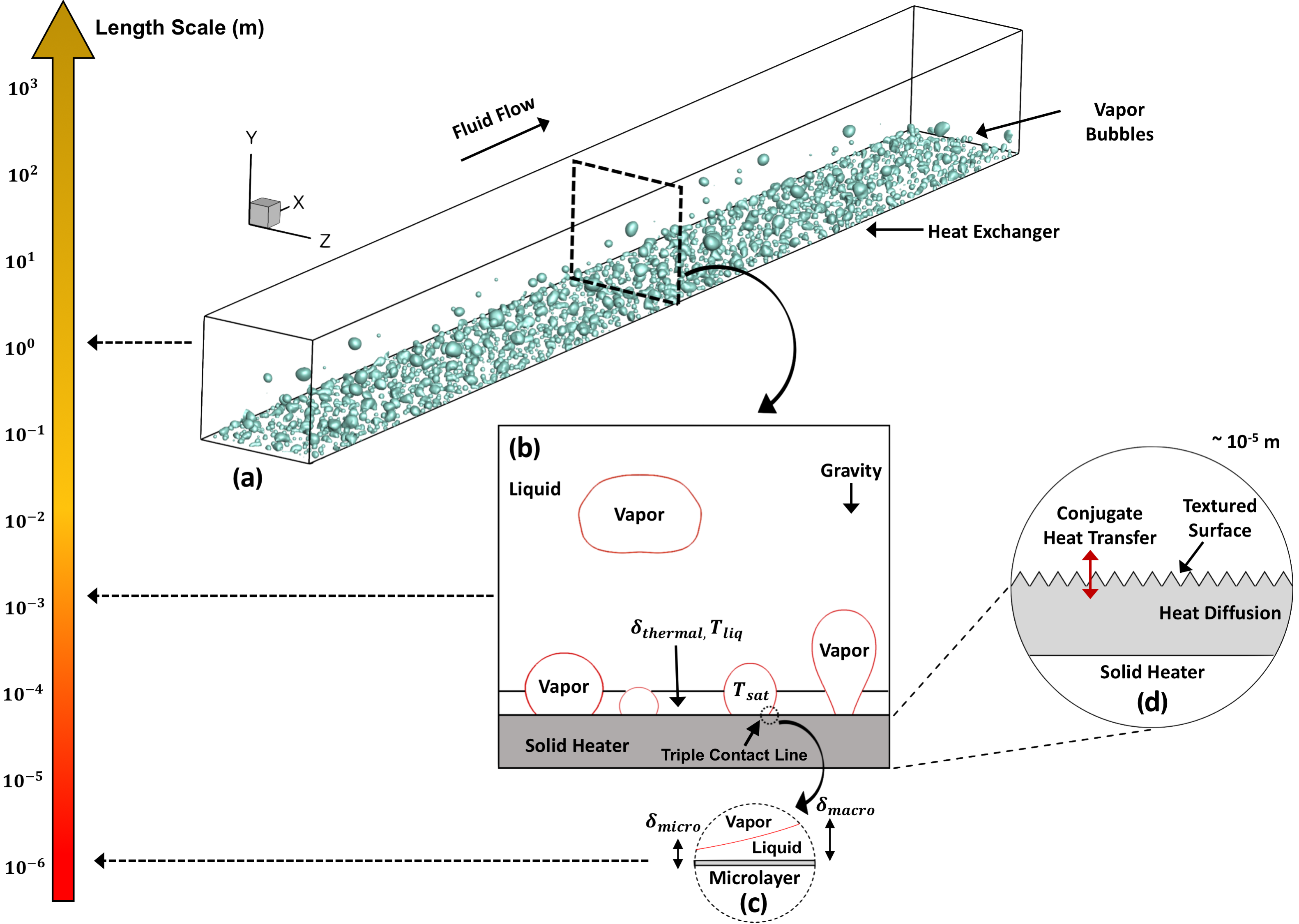}
\caption{An example of a target problem. (a) Flow boiling over a heat-exchanger (b) Schematic of the problem over a rough heater surface (c) Zoomed-in view of microlayer (d) Textured heater surface.\label{fig:scope}}
\end{center}
\end{figure}

These physics units are combined under the special unit \codetext{Simulation} to generate different application configurations. Figure \ref{fig:overview} provides an overview of application configurations built using a combination of different solvers discussed above. A broad classification of these applications can be described under fluid-solid interactions, fluid-fluid interactions, and scalar transport, typically involving heat transfer and/or species chemistry/evolution. Practical engineering simulations often utilize two or more of these units to account for interactions occurring at similar or different length and time scales. Flash-X brings all these capabilities together to create a powerful software framework that can be utilized for different types of fluid dynamics simulations. The details of numerical implementation that corresponding to the discussion in this article are documented in \cite{dhruv2023vortex,akash_phd_2021,DHRUV2019,DHRUV2021}.

Consider the problem of boiling in a heat exchanger which lies at the center of the Venn diagram in Fig \ref{fig:overview}. Boiling is the most efficient way of cooling machine components on earth, and is extensively used in internal combustion engines, industrial power plants, and nuclear reactors. Figure \ref{fig:scope}a shows a section of a liquid cooled heat-exchanger of length $\sim 1$ m, which is representative of the scale of industrial components. Due to high temperature on the heated surface the liquid undergoes phase change, enabling formation of vapor bubbles and leading to boiling (Figure~\ref{fig:scope}b). The process of evaporation occurs at the solid-liquid-gas triple contact line, near the heater surface, and within the thermal boundary layer of thickness, $\delta_{thermal} \sim 10^{-4}$ m, and microlayer of thickness, $\delta_{micro} \sim 10^{-6}$ m. The former consists of superheated liquid which by definition has temperature greater than the saturation temperature ($T_{liq}>T_{sat}$). The temperature of the superheated liquid, $T_{liq}$, is dependent on heat transfer through the heater wall, whereas the saturation temperature, $T_{sat}$, is based on fluid-vapor properties. An enlarged view is shown in Figure~\ref{fig:scope}c which highlights the microlayer which is comprised of liquid that gets trapped below the bubble base during rapid growth. The topology of the heater surface with the length scale of surface textures, $l ~ \sim 10^{-5}$m, also affects evaporation by exerting wall adhesion force on the bubbles and through conjugate heat transfer with the fluids (see Figure~\ref{fig:scope}d).
\begin{figure}[h]
\begin{center}
\includegraphics[width=0.85\textwidth]{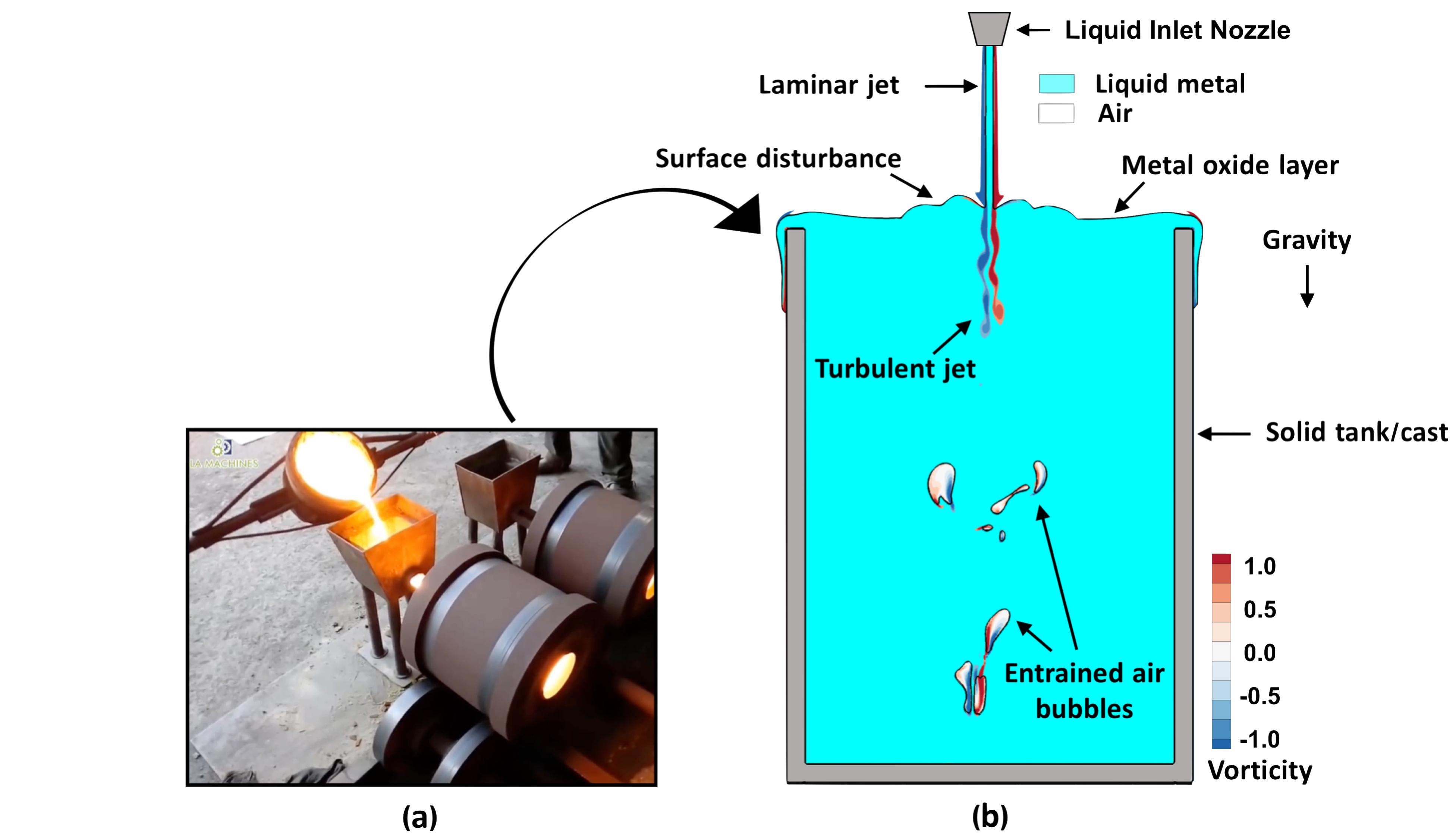}
\caption{(a) Demonstration of metal casting \url{https://www.youtube.com/@LAMachines} (b) Schematic of the problem.\label{fig:casting}}
\end{center}
\end{figure}

The disparity of scales that exist in the example problem of boiling has presented challenges in scientific understanding of the process. Experimental investigations undertaken in the past have been helpful in providing an understanding of marcoscale phenomena like scaling of average wall heat flux and bubble dynamics based on gravity and heater temperature \cite{KIM20023919}. It has also enabled calculation of quantitative data for growth and departure during single bubble nucleate boiling \cite{KIM2006208,Dhir2012}, and paved development of heuristic and data-driven models for nucleation site distribution and microlayer contribution \cite{Raj2012,URBANO20181128,UTAKA2013222}. However, fundamental knowledge relating to events associated with bubble dynamics (sliding, mergers, departure, etc.), and their relationship to fluid flow and heat transfer is still lacking, which can be acquired by running simulations.

This is the case with many problems described in Fig \ref{fig:overview}. The problem of impinging liquid jet during metal casting is another example where complexity is introduced through scale and physical dynamics of the process. Fig \ref{fig:casting} provides a schematic of the problem which shows a liquid metal jet falling on a pool of liquid metal due to gravity. The jet is laminar in air and becomes turbulent inside the pool as described by the vorticity contours in Fig \ref{fig:casting}b. Due to this sudden change in dynamics, disturbances are created at the air-liquid free surface which lead to entrainment of air bubbles. Furthermore, chemical reactions that occur at liquid-air interface (at free surface and near bubbles) lead to formation of metal-oxide impurity, which makes the metal brittle after solidification. The amount of metal-oxide formed during this process depends on nozzle design, shape of the liquid metal jet in air, and impact velocity at the free surface, whose relationship with one another is difficult to quantify experimentally \cite{griffin1991ladle}.

For the class of problems discussed above and others shown in Fig \ref{fig:overview}, numerical simulations become imperative for scientific discovery. They can provide high-fidelity spatio-temporal data, which can be used for data analysis and the discovery of new physics. The open-source development of Flash-X for modeling these problems serves as a significant incentive for the broader research community.
%
%
%
\section{Software Design and Composability} \label{sc: software-design}
\begin{figure}[h]
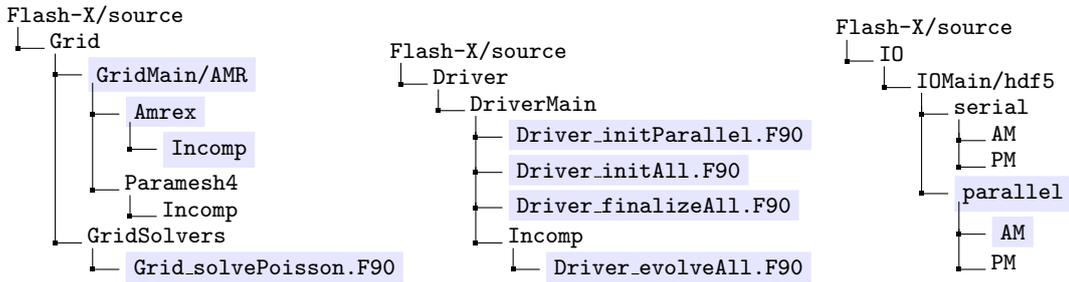

\begin{center}
\begin{subfigure}[b]{0.3\textwidth}
\mydirtree{%
.1 Flash-X/source.
.2 Grid.
.3 \highlight{GridMain/AMR}.
.4 \highlight{Amrex}.
.5 \highlight{Incomp}.
.4 Paramesh4.
.5 Incomp.
.3 GridSolvers.
.4 \highlight{Grid\_solvePoisson.F90}.
}
\end{subfigure}
\begin{subfigure}[b]{0.35\textwidth}
\mydirtree{%
.1 Flash-X/source.
.2 Driver.
.3 DriverMain.
.4 \highlight{Driver\_initParallel.F90}.
.4 \highlight{Driver\_initAll.F90}.
.4 \highlight{Driver\_finalizeAll.F90}.
.4 Incomp.
.5 \highlight{Driver\_evolveAll.F90}.
}
\end{subfigure}
\begin{subfigure}[b]{0.25\textwidth}
\mydirtree{%
.1 Flash-X/source.
.2 IO.
.3 IOMain/hdf5.
.4 serial.
.5 AM.
.5 PM.
.4 \highlight{parallel}.
.5 \highlight{AM}.
.5 PM.
}
\end{subfigure}
\end{center}
\caption{Organization of infrastructure units for Grid, Driver, and I/O. The highlighted boxes correspond to the units/source files required for simulation shown in Figure \ref{fig:scope}.}
\label{fig:source-infrastructure}
\end{figure}

In this section, we provide a more detailed description of the directory-based inheritance that allows for composability shown in Figure \ref{fig:overview}. In the simplest terms, directory-based inheritance implies that a directory tree starting from \codetext{Flash-X/source} to a target directory within a unit becomes a collection of node objects which inherit source code files and functionality from their parent node. To illustrate, refer to Figure \ref{fig:source-infrastructure}, which provides details for the infrastructure units--\codetext{Grid}, \codetext{Driver}, and \codetext{IO}. When an application requests a specific directory, such as \codetext{Grid/GridMain/AMR/Amrex/Incomp}, all the files along this directory tree are unpacked into \codetext{Flash-X/object} directory. The files in the child node replace or supersede the files from the parent node. 

Each node contains a \codetext{Config} file written in a Domain-Specific Language (DSL), parsed by a special \codetext{setuptool} to create \codetext{Flash-X/object} (an instance of the \codetext{Flash-X/source}). Links to source files under \codetext{Flash-X/object} can be compiled to build an executable for one of many simulations in Figure \ref{fig:overview}. Details of the DSL for \codetext{Config} files and the usage of \codetext{setuptool} have been extensively documented in the user's guide (\url{https://flash-x.org/pages/documentation}).

\codetext{Grid/GridMain/AMR} unit provides interface to two mutually exclusive AMR packages--AMReX \cite{AMReX_JOSS} and custom implementation of Paramesh \cite{Paramesh}, each containing special directories \codetext{Incomp} for customizations related to incompressible Navier-Stokes formulation. The \codetext{Grid/GridSolvers} unit provides a parallel directory structure to \codetext{Grid/GridMain} containing implementation for multigrid Poisson solvers. AMReX-based simulations use its native Poisson solver, while Paramesh-based simulations leverage linear solvers provided by HYPRE \cite{hypre}. The \codetext{Driver/DriverMain} unit, responsible for integrating different units and managing physics evolution, follows a similar structure and set of rules to \codetext{Grid/GridMain}. The file \codetext{Driver\_evolveAll.F90} under \codetext{Driver/DriverMain/Incomp} controls the evolution of multiphase fluid dynamics solution. Further details about this file will be discussed later, following the details of the \codetext{physics} and \codetext{Stencils} unit.

Finally, the \codetext{IO/IOMain/hdf5} unit provides an interface to the Hierarchical Data Format (HDF5) library for writing and reading data to and from the disk. Interfaces are available for both serial and parallel versions of the HDF5 library, tuned to the available AMR implementations--\codetext{AM} for AMReX and \codetext{PM} for Paramesh. Other infrastructure units like \codetext{Particles} and \codetext{RuntimeParameters} organize source code in a similar fashion. We strongly encourage reading the Flash-X user guide for the details (\url{https://flash-x.org/pages/documentation}).
\begin{figure}[h]
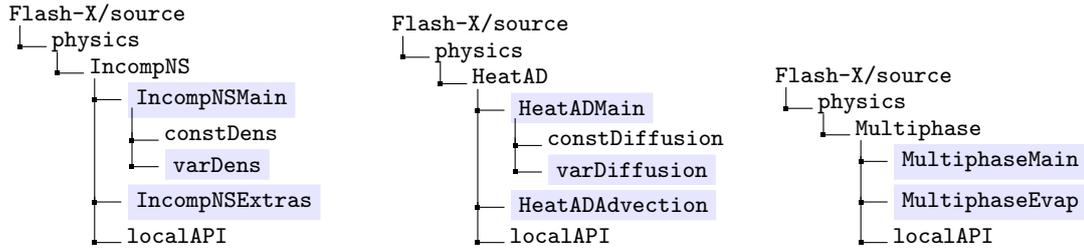

\begin{center}
\begin{subfigure}[b]{0.3\textwidth}
\mydirtree{%
.1 Flash-X/source.
.2 physics.
.3 IncompNS.
.4 \highlight{IncompNSMain}.
.5 constDens.
.5 \highlight{varDens}.
.4 \highlight{IncompNSExtras}.
.4 localAPI.
}
\end{subfigure}
\begin{subfigure}[b]{0.3\textwidth}
\mydirtree{%
.1 Flash-X/source.
.2 physics.
.3 HeatAD.
.4 \highlight{HeatADMain}.
.5 constDiffusion.
.5 \highlight{varDiffusion}.
.4 \highlight{HeatADAdvection}.
.4 localAPI.
}
\end{subfigure}
\begin{subfigure}[b]{0.3\textwidth}
\mydirtree{%
.1 Flash-X/source.
.2 physics.
.3 Multiphase.
.4 \highlight{MultiphaseMain}.
.4 \highlight{MultiphaseEvap}.
.4 localAPI.
}
\end{subfigure}
\end{center}
\caption{Organization of physics units for incompressible fluid dynamics, heat transfer, and multiphase equations. The highlighted boxes correspond to the units/source files required for simulation shown in Figure \ref{fig:scope}.}
\label{fig:source-physics}
\end{figure}

\begin{figure}[h]
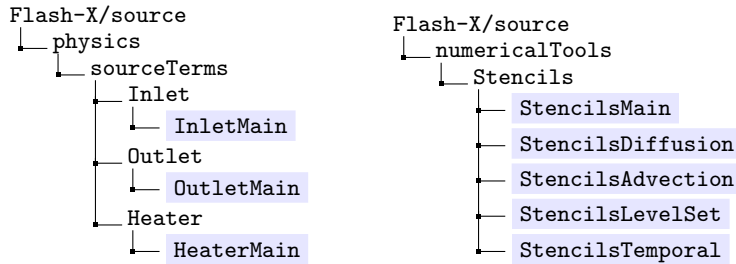

\begin{center}
\begin{subfigure}[b]{0.3\textwidth}
\mydirtree{%
.1 Flash-X/source.
.2 physics.
.3 sourceTerms.
.4 Inlet.
.5 \highlight{InletMain}.
.4 Outlet.
.5 \highlight{OutletMain}.
.4 Heater.
.5 \highlight{HeaterMain}.
}
\end{subfigure}
\begin{subfigure}[b]{0.4\textwidth}
\mydirtree{%
.1 Flash-X/source.
.2 numericalTools.
.3 Stencils.
.4 \highlight{StencilsMain}.
.4 \highlight{StencilsDiffusion}.
.4 \highlight{StencilsAdvection}.
.4 \highlight{StencilsLevelSet}.
.4 \highlight{StencilsTemporal}.
}
\end{subfigure}
\end{center}
\caption{Organization of units for source terms and stenciled operations. The highlighted boxes correspond to the units/source files required for simulation shown in Figure \ref{fig:scope}.}
\label{fig:source-stencils}
\end{figure}

Figure \ref{fig:source-physics} provides details of three \codetext{physics} units which model incompressible fluid dynamics, heat transfer, and multiphase interactions. The \codetext{physics/IncompNS/IncompNSMain} unit provides functionality for fractional-step project method for solving Navier-Stokes equations using either constant coefficient formulation (\codetext{constDens}) or variable coefficient formulation (\codetext{varDens}). The equations follow the style of Equation \ref{eq:transport} and leverage the design patterns by using functionality from \codetext{StencilsDiffusion}, \codetext{StencilsAdvection}, and \codetext{StencilsTemporal} from \codetext{numericalTool/Stencils} unit (see Figure \ref{fig:source-stencils} for details). The auxiliary unit \codetext{physics/IncompNS/IncompNSExtras} provides additional functionality to manage data interpolation from face-centers to cell-centers. This is useful for post processing and visualization.

The parallel units \codetext{physics/HeatAD/HeatADMain} and \codetext{physics/HeatAD/HeatADAdvection} offer implementations for heat diffusion and advection, respectively. The parallel design of this unit permits independent utilization for heat transfer problems that do not involve fluid flow. When using heat transfer with fluid flow, the advection unit will always be invoked. Similar to \codetext{physics/IncompNS/IncompNSMain}, these units leverage stencil design patterns based on their individual requirements. The sub-directories \codetext{constDiffusion} and \codetext{varDiffusion} correspond to constant and variable values for diffusion coefficient diffusion $K$ in Equation \ref{eq:transport}.

The \codetext{physics/Multiphase/MultiphaseMain} unit provides functionality for treating the level-set function $\phi$ that models the air-water (gas-liquid) interface in Figure \ref{fig:grid}. Currently, this unit has a hard dependency on the \codetext{varDens} configuration under \codetext{physics/IncompNS/IncompNSMain}, which is explicitly specified in its \codetext{Config} file. In addition to utilizing functionality from \codetext{StencilsAdvection} and \codetext{StencilsTemporal}, this unit also exploits design patterns for the level-function by using \codetext{StencilsLevelSet} to handle level-set specific operations for setting variable fluid properties and other relevant numerical procedures. Level-set operations are general and can be extended to other numerical algorithms that do not involve multiphase flows. Therefore, separating out this functionality provides a composable design for future work. 

Evaporation is modelled using \codetext{physics/Multiphase/MultiphaseEvap} unit which provides additional functionality to the \codetext{physics/Multiphase/MultiphaseMain} implementation. It is important to note that for any unit implementation, the main implementation will always be invoked before the auxiliary implementations. This unit has a hard dependency on \codetext{varDiffusion} configuration under \codetext{physics/HeatAD/HeatADMain}.

Figure \ref{fig:source-physics} also shows the \codetext{localAPI} for the physics units. \codetext{localAPI} exists under many other units and provides functionality that is private to the unit it belongs to. Public interface for a unit are named as, \codetext{Unit\_interfaceName}, and use a \codetext{"Unit\_"} prefix. The interfaces under \codetext{localAPI} use a different prefix (usually lower case) to denote that they should never be used in a unit outside their scope.
\begin{figure}[h]
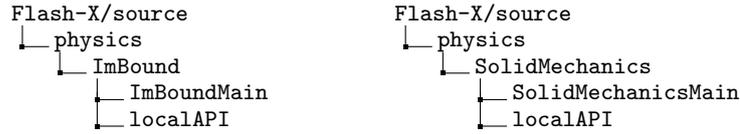

\begin{center}
\begin{subfigure}[b]{0.3\textwidth}
\mydirtree{%
.1 Flash-X/source.
.2 physics.
.3 ImBound.
.4 ImBoundMain.
.4 localAPI.
}
\end{subfigure}
\begin{subfigure}[b]{0.3\textwidth}
\mydirtree{%
.1 Flash-X/source.
.2 physics.
.3 SolidMechanics.
.4 SolidMechanicsMain.
.4 localAPI.
}
\end{subfigure}
\end{center}
\caption{Organization of immersed boundary and solid mechanics units for solid body, $\lambda$.}
\label{fig:source-imbound}
\end{figure}

The solid body $\lambda$ is primarily managed by \codetext{physics/SolidMechanics/SolidMechanicsMain}. This module utilizes the \codetext{Particles} unit to handle the Lagrangian data structure. The \codetext{SolidMechanics} unit manages the evolution of the body, influenced either by prescribed kinematics or fluid-induced forces.

The effect of body on the continuum flow is modelled using \codetext{physics/ImBound/ImBoundMain} which map the Lagrangian representation to a level-set function $\lambda_\phi$. This unit also leverages the level-set design patterns in \codetext{StencilsLevelSet}. The mapping uses an an Approximate Nearest Library (ANN) algorithm for efficient adaptation as the body moves within the domain. $\lambda_\phi$ effectively converts the Lagrangian reference frame to an Eulerian one, enabling the computation of $f$ in Equation \ref{eq:transport} to model the effects of the solid body on velocity, temperature, and pressure.

\begin{figure}[h]
\begin{center}
\includegraphics[width=0.65\textwidth]{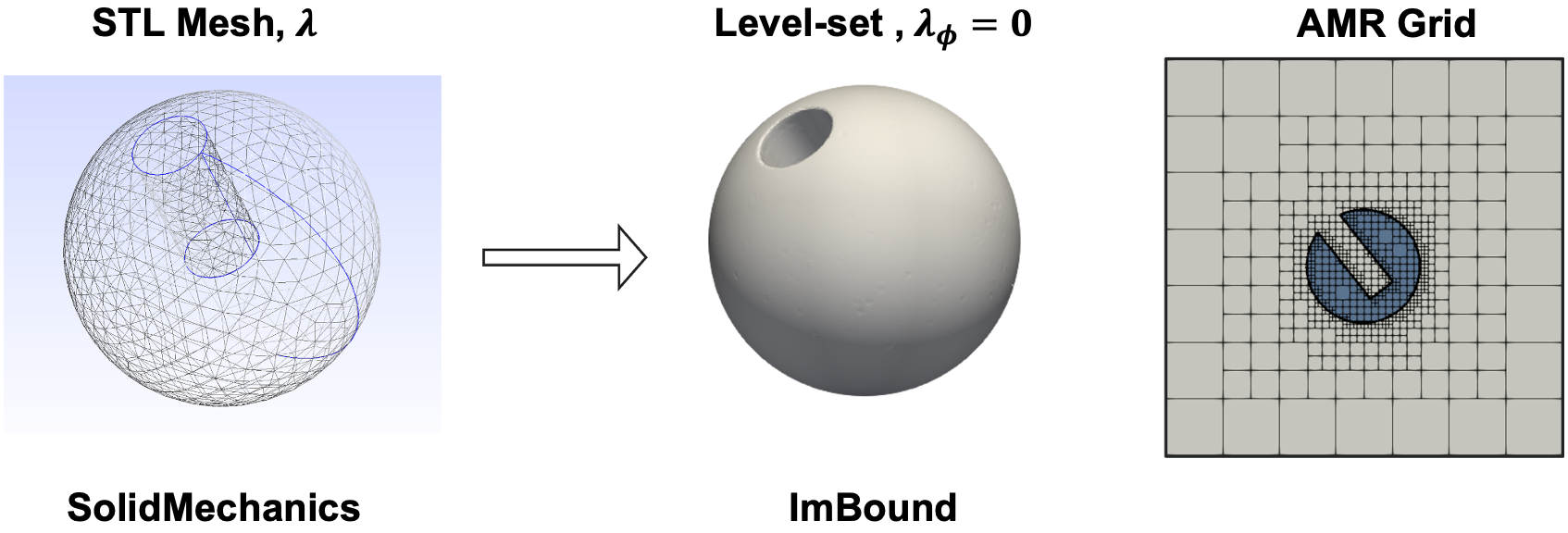}
\caption{Representation of the solid body $\lambda$ under two different units: \codetextsmall{SolidMechanics} unit uses Lagrangian reference frame and \codetextsmall{ImBound} maps it to a level set function $\lambda_\phi$.\label{fig:solid-body}}
\end{center}
\end{figure}

Figure \ref{fig:source-imbound} and \ref{fig:solid-body} provide details of the unit organization and the schematic of the Lagrangian-to-Eulerian mapping. The numerical details of interpolating and extrapolating fluid-body forces, implemented using a Moving Least Squares (MLS) scheme, are discussed in \cite{akash_phd_2021, Vanella2009, Vanella2010}.
\begin{figure}[h]
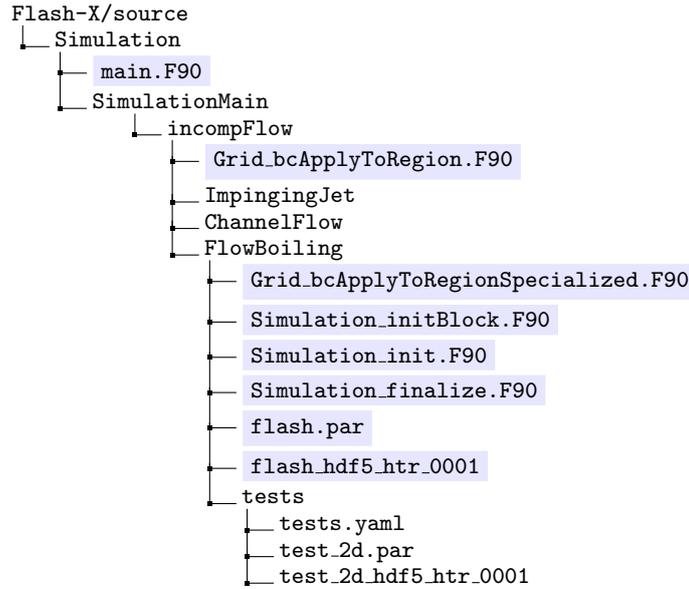

\begin{center}
\begin{subfigure}[b]{0.6\textwidth}
\mydirtree{%
.1 Flash-X/source.
.2 Simulation.
.3 \highlight{main.F90}.
.3 SimulationMain.
.5 incompFlow.
.6 \highlight{Grid\_bcApplyToRegion.F90}.
.6 ImpingingJet.
.6 ChannelFlow.
.6 FlowBoiling.
.7 \highlight{Grid\_bcApplyToRegionSpecialized.F90}.
.7 \highlight{Simulation\_initBlock.F90}.
.7 \highlight{Simulation\_init.F90}.
.7 \highlight{Simulation\_finalize.F90}.
.7 \highlight{flash.par}.
.7 \highlight{flash\_hdf5\_htr\_0001}.
.7 tests.
.8 tests.yaml.
.8 test\_2d.par.
.8 test\_2d\_hdf5\_htr\_0001.
}
\end{subfigure}
\end{center}
\caption{Organization of the simulation unit which contains different applications and use cases. The highlighted boxes correspond to the units/source files required for simulation shown in Figure \ref{fig:scope}.}
\label{fig:source-simulation}
\end{figure}

As mentioned earlier, the \codetext{Simulation} unit in Figure \ref{fig:source-overview} is a special unit where physics and infrastructure units are assembled to generate different applications. Figure \ref{fig:source-simulation} provides details for this unit. We would like to note that the actual \codetext{Simulation} under Flash-X provides a wide range of applications assembled using physics units that are not discussed in this article. To maintain focus on multiphase fluid dynamics applications, we intentionally left out references to other physics and simulation units, which have been extensively documented elsewhere \cite{COUCH2021102830}. 

The \codetext{main.F90} file under \codetext{Simulation} is the main program file for a Flash-X application. The contents of this file are shown in Figure \ref{fig:simulation-main}, which show calls to four top-level \codetext{Driver} interface procedures including \codetext{Driver\_evolveAll}, located under \codetext{Driver/DriverMain/Incomp} for fluid dynamics simulations. 
\begin{figure}[h]
\begin{minted}[
    frame=single,
    fontsize=\small,
  ]{fortran}
program Flashx
  use Driver_interface, ONLY : Driver_initParallel, Driver_initAll, &
                               Driver_evolveAll, Driver_finalizeAll
  implicit none
  call Driver_initParallel() ! located under Driver/DriverMain, intializes parallel environment
  call Driver_initAll() ! located under Driver/DriverMain, calls Unit_init for all units
  call Driver_evolveAll() ! located under Driver/DriverMain/Incomp
  call Driver_finalizeAll() ! located under Driver/DriverMain, call Unit_finalize for all units
end program Flashx
\end{minted}
\caption{Contents of \codetext{Simulation/main.F90}}
\label{fig:simulation-main}
\end{figure}

Different multiphase fluid dynamics applications are encoded further along the directory under \codetext{incompFlow} implementation of \codetext{Simulation/SimulationMain}, which shows three different applications, \codetext{FlowBoiling} (Figure \ref{fig:scope}), \codetext{ImpingingJet} (Figure \ref{fig:casting}), and a simple single-phase \codetext{ChannelFlow} simulation. The actual Flash-X source contains many other applications. To build the \codetext{Flash-X/object} directory for a given simulation, we can invoke the following \codetext{setuptool} call from \codetext{Flash-X} root directory,
\begin{minted}[
    frame=none,
    fontsize=\small,
  ]{bash}
  ./setup incompFlow/FlowBoiling -auto -maxblocks=100 -3d -nxb=16 -nyb=16 -nzb=16 \
                                       +amrex +incomp +parallelIO -site=hello
\end{minted}
which builds a three dimensional flow boiling simulation shown in Figure \ref{fig:scope} using files and directories highlighted in Figures \ref{fig:source-infrastructure} - \ref{fig:source-simulation}. The parameter file \codetext{flash.par} contains information for runtime parameters used to configure boundary conditions, domain size, AMR refinement levels, etc. The \codetext{flash\_hdf5\_htr\_0001} file contains HDF5 encoded information related to the heater configuration which is handled by \codetext{physics/sourceTerms/Heater}.

The corresponding snippet of \codetext{Driver\_evolveAll.F90} is shown in Figure \ref{fig:driver-evolve} which shows of different infrastructure and grid units come together to calculate predicted velocities for the fractional-step project method. The procedures \codetext{IncompNS\_advection} and \codetext{IncompNS\_predictor} are called each time step between \codetext{dr\_nBegin} and \codetext{dr\_nend} (the prefix \codetext{"dr\_"} indicates that variables are owned by the \codetext{Driver/DriverMain} unit). These procedures invoke design patterns in \codetext{StencilsAdvection} and \codetext{StencilsTemporal}, respectively. \codetext{IncompNS\_diffusion} uses a specialized design pattern for the \codetext{varDens} configuration, which is provided through its \codetext{localAPI} procedure \codetext{ins\_diffusion\_vardens}. Note that all the physics procedures operate on the data structure \codetext{tileDesc}, which represents a single or a collection of blocks ($\Omega$). \codetext{tileDesc} is allocated during loop over a specialized iterator which is managed through the \codetext{Grid} unit to allow for performance portability across different computing platforms. The details on the implementation of this performance portability layer are beyond the scope of the current article, and we aim to address them in a follow-up work.

An important design feature in Figure \ref{fig:driver-evolve} is the call to \codetext{Grid\_fillGuardCells}, which applies physical boundary conditions and performs halo exchange between AMR blocks. In FLASH, these calls were buried inside the physics units, making performance portability and optimization cumbersome. Within Flash-X, these calls are exposed to the top-level Driver interface, leading to a cleaner software design.
\begin{figure}[h]
\begin{minted}[
    frame=single,
    fontsize=\small,
  ]{fortran}
subroutine Driver_evolveAll()
  ! use statements referencing prodecures from different units that are needed...
  implicit none
  ! variable declarations and book-keeping...
  do dr_nstep = dr_nBegin, dr_nend ! Evolution loop over steps
    ! preceeding physics related calls...
    ! ...
    call Grid_fillGuardCells() ! Perform halo-exchange and apply boundary conditions
  
    call Grid_getTileIterator(itor, nodetype=LEAF) ! Get the tile iterator for LEAF blocks
    do while (itor%isValid()) ! Loop over iterator
      call itor%currentTile(tileDesc) ! Get tile descriptor from the iterator
      call IncompNS_advection(tileDesc) ! uses procedures from StencilsAdvection
      call IncompNS_diffusion(tileDesc) ! uses procedures from IncompNS/localAPI
      call IncompNS_predictor(tileDesc, dr_dt) ! uses procedures from StencilsTemporal
      call Multiphase_velForcing(tileDesc, dr_dt) ! dr_dt is time-step
      call itor%next() ! Increment the iterator
    end do
    call Grid_releaseTileIterator(itor) ! Release the iterator
    ! ...
    ! succeding physics related calls...
  end do
  ! book-keeping and clean up...
end subroutine Driver_evolveAll
\end{minted}
\caption{Snippet of \codetext{Driver/DriverMain/Incomp/Driver\_evolveAll.F90} to solve the predictor step for the fractional-step projectio method \cite{dhruv2023vortex}.}
\label{fig:driver-evolve}
\end{figure}

\section{Testing Framework and Scientific Reproduciblity} \label{sc: testing-reproducibility}
To ensure the accuracy and sustainability of the codebase, the Flash-X team has dedicated a significant amount of time to designing tools and procedures for testing, maintenance, and reproducibility. Two command line tools, in particular, have been developed:
\begin{itemize}
    \item \codetext{flashxtest}: A testing framework enabling uniform testing practices across different computing platforms, research groups, and individual developers \url{https://github.com/Flash-X/Flash-X-Test}.
    \item \codetext{jobrunner}: A utility to build and manage laboratory notebooks for systematic curation of execution environments, simulation data, and post-processing scripts and procedures. This tool facilitates reproducibility and collaboration efforts \url{https://github.com/Lab-Notebooks/Jobrunner}.
\end{itemize}
\begin{figure}[h]
\begin{center}
\includegraphics[width=0.85\textwidth]{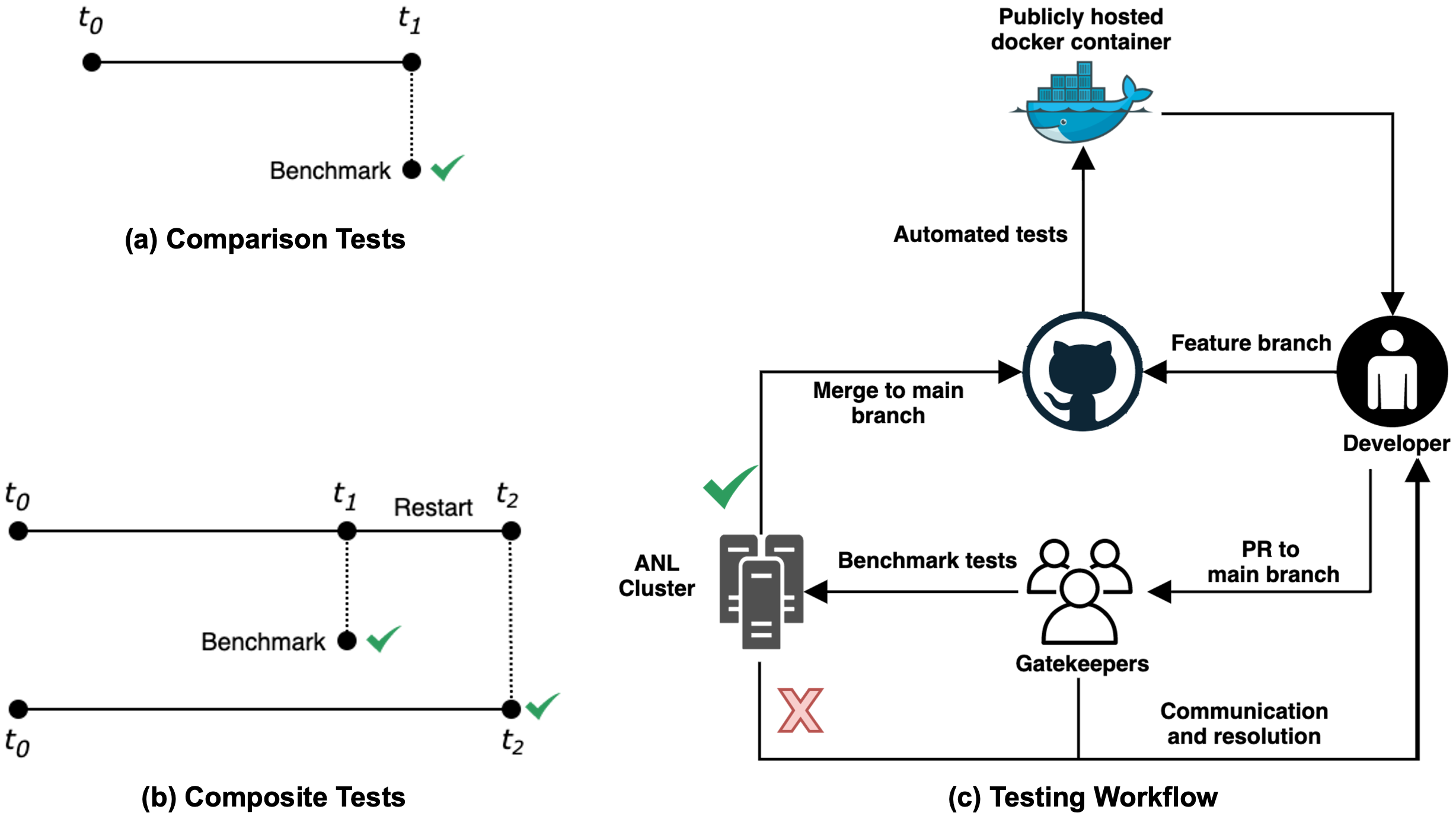}
\caption{Overview of testing framework: (a) Comparison tests that run a simulation from time $t_0$ to $t_1$ and compare the results against an approved benchmark, (b) Composite tests involve restarting simulations from $t_1$ to $t_2$ to ensure that restarts are transparent, and (c) Testing workflow for on different \codetextsmall{sites} -- a cluster hosted at Argonne National Laboratory (ANL) and a public docker container. Gatekeepers manage testing on the ANL cluster and maintain control over the quality of the source code in the main branch of the GitHub repository, while developers leverage the public docker container to perform unrestricted testing of their code on a feature branch \cite{dhruv2023framework}.}
\label{fig:testing}
\end{center}
\end{figure}

Figure \ref{fig:testing} provides an overview of the testing framework. Flash-X test specifications are essentially shorter version of applications located under \codetext{Simulation/SimulationMain} unit. See Figure \ref{fig:source-simulation} for example. The file \codetext{tests/tests.yaml} consists test specifications for \codetext{FlowBoiling} application under \codetext{Simulation/SimulationMain/incompFlow}. Figure \ref{fig:tests-yaml} provides an example of the contents within this file, which contains information related to setup options, parfile, and auxiliary file transfers required to run the \codetext{Comparison/incompFlow/FlowBoiling/2d/AMReX} test.
\begin{figure}[h]
\begin{minted}[
    frame=single,
    fontsize=\small,
  ]{yaml}
Comparison/incompFlow/FlowBoiling/2d/AMReX:
  setupOptions: -auto -maxblocks=100 +amrex +parallelIO +incomp -2d -nxb=16 -nyb=16
  parfiles: test_2d.par
  transfers: source/Simulation/SimulationMain/incompFlow/FlowBoiling/tests/test_2d_hdf5_htr_0001
\end{minted}
\caption{Example contents of \codetext{Simulation/SimulationMain/incompFlow/FlowBoiling/tests/tests.yaml}.}
\label{fig:tests-yaml}
\end{figure}

Flash-X supports unit tests and regression tests. Regression tests are comparison or composite tests as shown in Fig \ref{fig:testing}a and b. Comparison tests are based on the availability of an inspected and approved benchmark to compare the output generated by the application instance run as a test of correctness. Composite tests include two comparison tests, one in which the application starts from initial conditions and runs to a time $t_1$ and another in which the application starts from the checkpoint at time $t_1$ and runs to a time $t_2$. Therefore, each composite test requires two benchmarks. These two types of regression tests capture the real world need for simulations to reproduce correct data over time, and also perform transparent restarts when running production simulation on a HPC cluster where allocation time may not match the required simulation time. Testing against approved benchmarks is performed using Serial Flash Output Comparison Utility (SFOCU), available under \codetext{Flash-X/tools}.

Flash-X has brought together developers from different scientific domains, necessitating the need to employ uniform testing practices among different domains and research groups. This is facilitated using test suites, which represent collections of tests covered on a \codetext{site} (a HPC platform or a cloud architecture). Following is an example entry for a suite file which follows the style of Flash-X \codetext{setup} command,
\begin{minted}[
    frame=none,
    fontsize=\small,
  ]{bash}
  incompFlow/FlowBoiling -t "Comparison/incompFlow/FlowBoiling/2d/AMReX" -np 4 -cbase <yyyy-mm-dd>
\end{minted}

This command utilizes the test specifications provided in Figure~\ref{fig:tests-yaml}. It executes a test with four MPI processes and compares its results against the benchmark established on the date \codetext{<yyyy-mm-dd>}. These tests are managed and deployed using \codetext{flashxtest} which is extensively documented in \cite{dhruv2023framework}.

Figure~\ref{fig:testing}c provides an overview of the complete testing workflow. Individual developers, seeking an efficient process for setting up automated tests when implementing new features, employ a public Docker container to conduct unrestricted testing on their feature branches. In contrast, Flash-X gatekeepers, responsible for ensuring the quality and accuracy of scientific results, follow a more restrictive workflow with well-defined testing protocols. When individual developers create a pull request to the main branch, gatekeepers initiate testing on the Argonne National Laboratory cluster, utilizing a private and protected test suite. Test failures necessitate manual intervention, communication, and resolution between gatekeepers and developers before any new updates are merged into the main branch.

\begin{figure}[h]
\begin{center}
\includegraphics[width=0.85\textwidth]{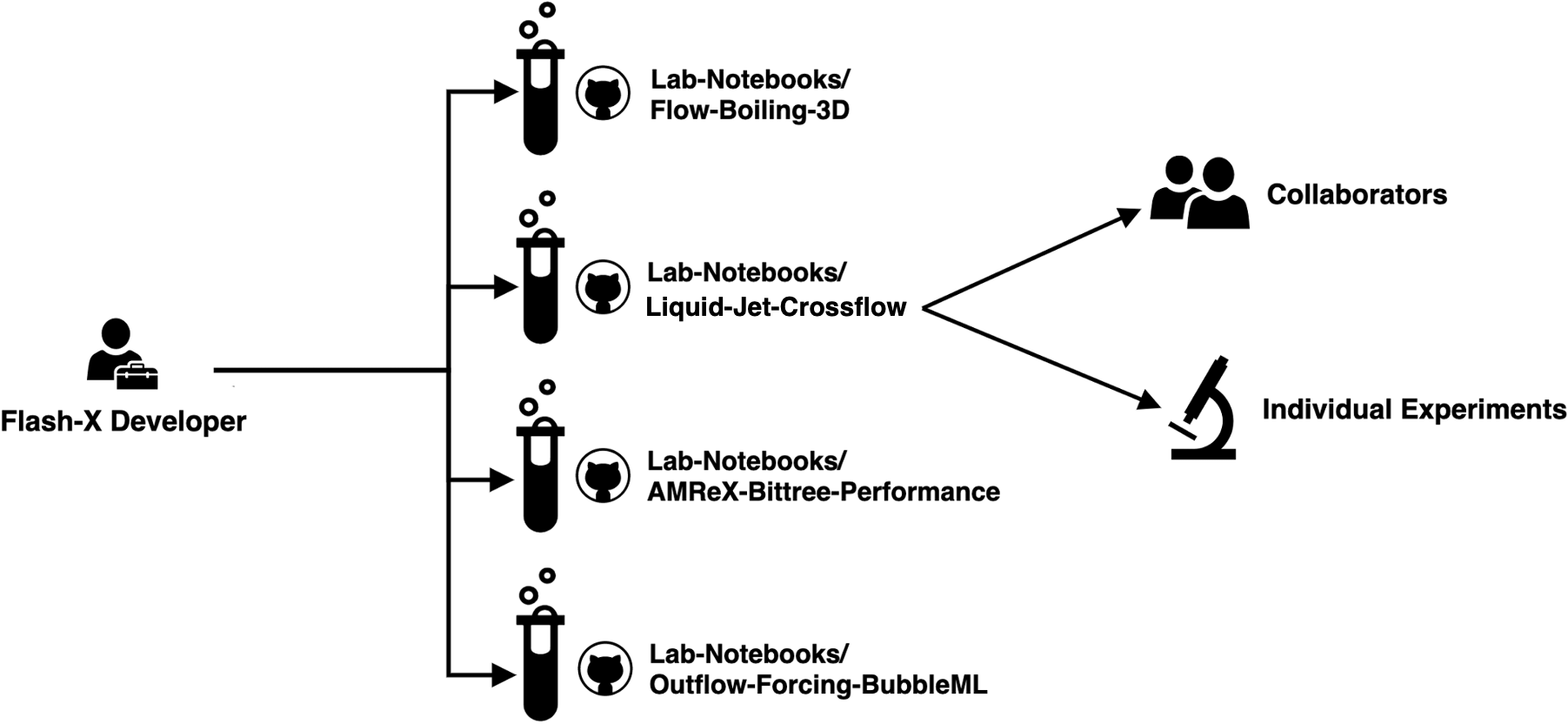}
\caption{A common scenario in computational sciences where a developer has to conduct individual experiments and collaborate with other researchers who may not be experts in using their software. Developers can create lab notebooks for different experiments related to performance and physics related studies and use them to document their investigation and collaborate with others. Copyright images from our IEEE publication \cite{Dhruv_Jobrunner_2023}.}
\label{fig:reproducibility}
\end{center}
\end{figure}

Scientific experiments carried out with Flash-X are meticulously cataloged and documented in individual repositories accessible via \url{https://github.com/Lab-Notebooks}. This approach enables collaborators and new developers to initiate their work with validated and well-documented studies, facilitating a smoother transition into the intricacies of the Flash-X codebase. Figure \ref{fig:reproducibility} showcases various laboratory notebooks crafted by a Flash-X developer, which serve as valuable resources for conducting individual experiments or for collaborative research endeavors. 

Laboratory notebooks capture the execution environment, software configurations, and other scientific decisions and leverage a directory-based inheritance similar to Flash-X to conduct and archive studies using \codetext{jobrunner}. A detailed documentation of this process can be found in \cite{Dhruv_Jobrunner_2023}.

The \codetext{jobrunner} utility also ensures the organization of runtime parameters. It achieves this by utilizing TOML configuration files as an interface to generate application-specific \codetext{flash.par} files. An illustrative snippet is provided in Figure \ref{fig:flash-toml}, showcasing the use of a nested implementation of TOML keys to declare runtime parameters under the respective units managing them.
\begin{figure}[h]
\begin{minted}[
    frame=single,
    fontsize=\small,
  ]{toml}
[Grid]
  xmin = 0.0 # Domain lower bound
  xmax = 1.0 # Domain upper bound
  xl_boundary_type = "inflow_ins" # Boundary condition at lower bound
  xr_boundary_type = "outflow_ins" # Boundary condition at upper bound
  nblockx = 10 # Number of blocks

[IncompNS]
  ins_invReynolds = 0.0001 # Inverse Reynolds number
  ins_gravY = -0.01 # Inverse square-root of Froude number

[Outlet]
  out_buffer = 1.0 # Outlet parameters

[Multiphase]
  mph_rhoGas = 0.001 # Ratio of gas to liquid density
  mph_muGas = 0.02 # Ratio of gas to liquid viscosity
  mph_invWeber = 0.004 # Inverse Weber number

[Driver]
  nend = 1000 # Total evolution steps
  tmax = 10 # Maximum simulation time

[IO]
  basenm = "INS_Simulation_" # Basename for HDF5 IO files
  plot_var_1 = "pres" # Plotfile variable
\end{minted}
\caption{Snippet of a TOML configuration file used by \codetext{jobrunner} to generate a parfile \url{https://github.com/Lab-Notebooks/Liquid-Jet-Crossflow}.}
\label{fig:flash-toml}
\end{figure}

Assigning runtime parameters under unit-specific keys enhances implicit documentation, thereby improving scientific reproducibility. Additionally, this practice serves as a foundational step in understanding the design of the Flash-X architecture. When a user or developer modifies a parameter, they can readily identify the specific unit they are interfacing with. This knowledge proves invaluable for resolving bugs, incorporating new functionality, or gaining deeper insights into the codebase.
%

%% file: Tex/Conclusion.tex
%
%
\section{Conclusion} \label{sc: conclusion}
In this article, we present a comprehensive overview of the infrastructure and physics capabilities within Flash-X. Our exploration includes the identification of various data structures and design patterns inherent in the mathematical framework of a computational system. We provide detailed insights into how Flash-X strategically leverages these structures and patterns to construct a robust software architecture specifically tailored for multiphase fluid dynamics simulations.

We proceed to demonstrate how this resulting architecture achieves composability within Flash-X, showcasing its adaptability in configuring diverse numerical simulations. These simulations encompass solid-fluid interactions, fluid-fluid dynamics featuring heat transfer, and phase transition phenomena.

We extend our discussion to encompass broader system design considerations, emphasizing the critical facets of software maintenance and reproducibility. We provide insight into established tools and practices that have evolved over the years in the development of Flash-X. Our intention is to not only showcase the current state of the software but also to contribute valuable knowledge to the community regarding sustainable design practices in scientific computing. 

This article is intended to serve as a guideline for Flash-X developers and the broader research community on the sustainable and modular design of scientific computing software. Additionally, we touch on aspects of performance portability, which we plan to elaborate on in a follow-up work.

\bigskip\noindent
\textbf{Acknowledgements:}
The author expresses gratitude to Dr. Anshu Dubey (Argonne National Laboratory), Dr. Klaus Weide (University of Chicago), and Jared O'Neal (Argonne National Laboratory) for their invaluable guidance and support in developing the Flash-X software design for multiphase fluid dynamics simulations. They have also been co-authors on other Flash-X related articles and proceedings. Additionally, sincere appreciation is extended to Dr. Elias Balaras (George Washington University), Dr. Marcos Vanella (National Institute of Standards and Technology), Keegan Delaney (Dsc. George Washington University), and Dr. Shizao Wang for their significant contributions to FLASH, which laid the foundation for the developments discussed in this article.

The material is based upon work supported by Laboratory Directed Research and Development (LDRD) funding from Argonne National Laboratory, provided by the Director, Office of Science, of the U.S. Department of Energy under contract DE-AC02-06CH11357 and the Exascale Computing Project (17-SC-20-SC), a collaborative effort of the US Department of Energy Office of Science and the National Nuclear Security Administration.

We also acknowledge that The City of Chicago is located on land that is and has 
long been a center for Native peoples. The area is the traditional homelands of 
the Anishinaabe, or the Council of the Three Fires: the Ojibwe, Odawa, and Potawatomi Nations. 
Many other Nations consider this area their traditional homeland, including the Myaamia, Ho-Chunk, Menominee, Sac and Fox, Peoria, Kaskaskia, Wea, Kickapoo, and Mascouten.

The submitted manuscript has been created by UChicago Argonne, LLC,
operator of Argonne National Laboratory (“Argonne”). Argonne, a
U.S. Department of Energy Office of Science laboratory, is operated
under Contract No. DE-AC02-06CH11357. The U.S. Government retains for
itself, and others acting on its behalf, a paid-up nonexclusive,
irrevocable worldwide license in said article to reproduce, prepare
derivative works, distribute copies to the public, and perform
publicly and display publicly, by or on behalf of the Government.  The
Department of Energy will provide public access to these results of
federally sponsored research in accordance with the DOE Public Access
Plan. http://energy.gov/downloads/doe-public-access-plan.

During the preparation of this work the author(s) used ChatGPT in order to clean up language in the manuscript. After using this tool/service, the author(s) reviewed and edited the content as needed and take(s) full responsibility for the content of the publication.